\documentclass[11pt,a4paper]{article}
\usepackage{amsmath,amssymb,graphicx,epsfig,cite,color,hyperref}
\usepackage[left=2.7cm,right=2.7cm,top=2.5cm,bottom=2.5cm]{geometry}
\usepackage{subfigure}
\usepackage{xcolor}

\begin{document}



\begin{center}
{\Large \bf The impact of the intrinsic charm quark content of the proton on differential $\gamma+c$ cross section}
\vspace{.7cm}

S. Rostami$^{1}$, A. Khorramian$^{1,2}$, A. Aleedaneshvar$^{1}$

\vspace{.3cm}
{\it ~$^1$ Physics Department, Semnan University, Semnan, Iran,\\
~$^2$  School of Particles and Accelerators, Institute for
Research in Fundamental Sciences (IPM), P.O.Box 19395-5531, Tehran,
Iran \\
}
\end{center}

\vspace{0.1cm}

\begin{center}
  {\bf \large Abstract}
\end{center}
We present a comparative analysis of the impact of 
the non-perturbative intrinsic charm quark content of the proton
  on differential
cross section of $\gamma +c $-jet 
 in $ pp $ and $ p\bar{p} $ collisions, for the kinematic regions
 that are sensitive to this contribution. We discuss
 the $ Q^2 $ evolution of intrinsic quark
distributions at the next-to-leading
order (NLO) and present a code which provide
these distributions as a function of  $  x $ and  $ Q^2 $ for any 
arbitrary Fock state probability. For the $ p\bar{p} $ collisions at the Tevatron,
 the results are compared with  the recent experimental data of D0
 at $ \sqrt{s} =1.96$~TeV and also  predictions
for $ pp $ collisions at $ \sqrt{s} =8$~TeV and $ \sqrt{s} =13$~TeV for the LHC.


\clearpage

\section{Introduction}\label{sec:int}
The parton distribution functions (PDFs) are essential
to make precise prediction for the standard model (SM)
processes at hadron colliders such as $ p\bar{p} $  and $ pp $ scattering
at Tevatron and LHC,  respectively. Actually, the cross section 
of hadron-hadron scattering is explained  in terms of 
parton-parton scattering which is the convolution of 
PDFs and partonic cross section.
One of the important properties of these non-perturbative PDFs
is that they are universal, i.e., they are the same in all kinds of processes.
Precise knowledge of these PDFs  describing the proton's quark 
and gluon content is very important to test the SM and to search for New Physics.

The parton distribution  $ f_i(x,Q^2)$ of the proton, is the number density
of partons of flavour $i$ carrying a momentum fraction $ x $ at energy scale $ Q^2 $.
Because the PDFs are non-perturbative objects, they
cannot be determined directly from the first principles in
QCD and have to be fixed by experimental information.
The procedure is  parametrizing the $ x $ dependence of
PDFs at a scale $ Q_0^2 $, where $ Q_0^2 $ has to be lie in
the perturbative regime such that the  known Dokshitzer-Gribov-Lipatov-Altarelli-Parisi (DGLAP) evolution equations 
 are applicable \cite{Altarelli:1977zs,Gribov:1972ri,Dokshitzer:1977sg}.

According to DGLAP evolution equations, we   are able to obtain 
the shape of the PDFs by fitting to the available data from experimental observables.
In recent years, several theoretical groups have
extracted PDFs by doing a QCD global analysis \cite{Harland-Lang:2014zoa,CTEQ,MSTW,NNPDF,DelDebbio:2007ee,Bourrely:2015kla,ABM,JR,KKT,Kusina:2014wwa,Arbabifar:2013tma,
Khorramian:2010qa}.


The QCD physics of heavy quarks in the proton is one of the most important purposes for 
the Electron-Ion Collider (EIC), and it has many important consequences for high energy colliders, including the 
Large Hadron Electron Collider (LHeC) at CERN. Recently a number of important processes, including heavy quarks production which are sensitive to
charm and bottom quark distributions are presented \cite{Abazov:2012ea,D0, D01,Abazov:2014hoa}. 
Indeed, Fock states of the proton wave function with five quarks such as
$uudq\bar q$ where $q=u, d, s$ and $c, b$ has been  considerably interested in 
recent years \cite{Lyonnet:2015dca, Lyonnet:2015sda, Brodsky:2015fna, 
Brodsky:2015uwa, Lansberg:2015lva,Ball:2015tna,vafaee,rostami, Speth98-1, Kumano98-2, Vogt00-3, Garvey01-4, 
Chang11-5, Peng14-6, Jimenez-Delgado:2014zga}.  In this regards, many articles have studied
non-perturbative ``intrinsic'' sea quark components in
addition to the commonly perturbative ``extrinsic'' ones in
the nucleon wave function, which the first time was suggested
by Brodsky, Hoyer, Peterson, and Sakai (BHPS)
in 1980 \cite{BHPS1} (see Ref. \cite{Brodsky:2015fna}
for a recent review).  

There are remarkable differences between the extrinsic
and intrinsic sea quarks. The extrinsic
sea quarks arise in the proton perturbatively
through the splitting of gluons into quark-antiquark
pairs in the DGLAP  $ Q^2 $ evolution
and produce more and more when the $ Q^2 $
scale increases. Meanwhile extrinsic sea quarks
dominate at very low parton momentum
fraction $  x $ and so have a ``sealike'' characteristics.
 In contrast, the intrinsic sea quarks arise
through the non-perturbative fluctuations of the nucleon state
to five-quark states or virtual meson-baryon states in Meson Cloud Model (MCM) framework \cite{Thomas,Kumano98-2}
in the light-cone
Fock space picture \cite{Brodsky1}. They exist over a time
scale which is independent of any probe
momentum transfer (infinite momentum frame).
Moreover, the intrinsic sea quarks behave as
 ``valencelike'' quarks and then their distributions
 peak at relatively large $  x $.

In addition to the BHPS, there have also been
a number of theoretical calculations
 to describe the intrinsic charm (IC) distribution
 in the light cone framework. However, 
 one can find a review of these models in Refs. \cite{Pumplin1,Hobbs1}.
  For example, the study of EMC charm leptoproduction
  data  done by Harris, Smith and Vogt \cite{Harris}
  indicated that an IC component with 0.86$ \pm $0.6\%
  probability  can be present  in the nucleon. Also the CTEQ collaboration \cite{pump,cteq66,Dulat}  studied 
  the magnitude of the probability for intrinsic
  charm state within a global analysis of PDFs considering
  a wide range of the hard-scattering data. They showed that
   the probability for IC can be 2-3 times larger than the predicted value without
  any inconsistency with the experimental data. 
      Recently, two global analyses about the importance of
intrinsic charm have been performed.
The first one, by the CTEQ collaboration \cite{Dulat}, follows their previous work 
and the second one, by Jimenez-Delgado {\it et al.} 
\cite{Jimenez-Delgado:2014zga} which using looser kinematic cuts
for including low-$ Q $ and high-$x$ data.
  Also in Refs.~\cite{Bednyakov,Bednyakov1} 
  searches for intrinsic charm component of the proton at the LHC 
  are presented. In this case, having a code which can extract 
  the intrinsic heavy densities for any
  arbitrary intrinsic quark probability would be valuable.

As mentioned before, some processes
are sensitive to charm quark distributions
in the large $ x $ region. For instance,
the produced charm quark in deep inelastic
lepton-proton scattering $ l p \rightarrow l^\prime c X $, 
which has been performed by
the European Muon Collaboration (EMC) experiment \cite{EMC}.
As another example, the  $ J/\psi  $  hadroproduction
at high $ x $  which was observed in  $ pA $ and $ \pi A  $ collision
 by  NA3
at CERN \cite{NA3} and E866 at FNAL \cite{FNAL}.
Likewise, the inclusive production of
charmed hadrons at hadron colliders
at large $ x $  is a good laboratory to
investigate the role of intrinsic charm in the proton.
For example, the production of charmed hadrons in
$ pp \rightarrow D X $
and
$ pp \rightarrow \Lambda_c X $
observed at ISR \cite{ISR} and at Fermilab \cite{Fermi1,Fermi2}.
In addition to the above mentioned processes,
the results of prompt photon production in association
with a charm quark at hadron colliders
($ pp(\bar{p}) \rightarrow \gamma +c $-jet)  \cite{D0,Bednyakov}
 are dependent  on the charm
quark distribution.  The contribution of
the charm quark at large $ x $ can be studied also in the $ c $-jet
production accompanied by vector bosons $ Z,W^{\pm} $ \cite{D01}.

In the present study, we perform the $ Q^2 $ evolution of intrinsic quark
distributions and present a code providing
these distributions at any $  x $ and  $ Q^2 $
values for any arbitrary Fock state probability. Besides we present a comparative analysis of
IC contribution in the
proton,
 using the  production of $\gamma +c $-jet
 in $ pp $ and $ p\bar{p} $ collisions. Particular attention is paid to calculate the differential $ \gamma+c $-jet cross section.
 We can compare our results with  the recent experimental data of D0
 at $ \sqrt{s} =1.96$~TeV at the Tevatron. Also we present some predictions
 for $ pp $ collisions at $ \sqrt{s} =8$~TeV and $ \sqrt{s} =13$~TeV for the LHC. The outline of this paper is as follows.
In section \ref{sec:cone}   we briefly review
 the light-cone model for intrinsic charm
 introduced by BHPS. The
 $ Q^2 $-evolution of intrinsic quark distribution is presented
 in Sec.~\ref{sec:evo}, where we compare the evolution of
 extrinsic and intrinsic  charm  distributions.
 In Sec.~\ref{sec:pho} we present our  NLO predictions
for production of $ \gamma+c $-jet  in $ pp $ and $ p\bar{p} $ collisions.
Finally, in Sec.~\ref{sam}
we will summarize our results.

\section{The light-cone picture of the proton and the BHPS model}\label{sec:cone}
The light cone formalism allows   a proton to exist in
various Fock configurations \cite{lightcone}.
Therefore the wave function of the proton consists of
$ \vert uud\rangle $ distribution plus Fock states of $n$-particle given by
number \cite{Brodsky}
\begin{eqnarray}
\vert p \rangle &=& \psi _{3q/p}(\vec{k}_{\perp i},x_i) \vert uud  \rangle
 +\psi _{3qg/p}(\vec{k}_{\perp i},x_i) \vert uudg \rangle \nonumber \\
 &+&\psi _{5q/p}(\vec{k}_{\perp i},x_i)\vert uudq\bar{q} \rangle+\ldots~.
\label{F1}
\end{eqnarray}
In the light cone framework for the wave function
amplitudes of Fock components we have
\begin{equation}
\psi_{n/p}(\vec{k}_{\perp i},x_i)\propto \dfrac{1}{M^2-\sum_{i=1}^n\left(\dfrac{m^2_i+k_{\perp i}^2}{x_i}\right)},
\end{equation}
where $ M $ is the mass of the proton and $ m_i $ and
$ k_{\perp i} $ are the mass and transverse momentum
of parton $ i $ in the Fock state, respectively and  $ x_i$ is the
momentum fraction  carrying by parton $ i $
  that the momentum conservation is satisfied as follows
 \begin{eqnarray}
\sum _{i=1}^n \vec{k}_{\perp i}=\vec{0}_{\perp},\,\,\,\, \sum _{i=1}^n x_i=1,
\label{E2}
\end{eqnarray}
and the momentum distributions are determined
from integrating the square of the wave function.

The possible existence of a five-quark Fock component
$ \vert uudq\bar{q}\rangle $ in the wave function of the proton for
the first time presented by
Brodsky, Hoyer, Peterson, and Sakai (BHPS) in 1980.
According to the BHPS model the probability
distribution for the five-quark state assuming
that the effect of transverse momentum is negligible, can be written as \cite{BHPS1}
 \begin{eqnarray}\label{E3}
P(x_1,...,x_5) & \! =  \! &  {\cal N} \delta (1-\sum _{i=1}^5 x_i)[M^2-\sum _{i=1}^5 \dfrac{m_i^2}{x_i}]^{-2}\nonumber\\
 & \! =  \! & {\cal N}  \delta (1-\sum _{i=1}^5 x_i)
    \Big[ M^2-\dfrac{m_u^2}{x_1}-\dfrac{m_u^2}{x_2}-\dfrac{m_d^2}{x_3}-\dfrac{m_q^2}{x_4}-\dfrac{m_{\bar{q}}^2}{x_5}\Big]^{-2},\nonumber\\
\end{eqnarray}
where as mentioned above, $m_i $ is the mass of the parton $ i $ in the Fock state and $ x_i$
is the momentum fraction carried by it.
In the above equation, $  {\cal N}$ is the normalization factor and can be determined from
\begin{equation}\label{p5}
{\cal P}_5^{q\bar{q}}= \int_0^1 dx_1...dx_5 P(x_1,...,x_5),
\end{equation}
where ${\cal P}_5^{q\bar{q}}  $ is the $ \vert uudq\bar{q}\rangle $
Fock state probability.
For the case that $ q $ is a heavy quark, namely $ c $ or  $ b $
(which we denote here with $ Q $),    BHPS
assumed   the light quark and proton masses are negligible
 compared to the heavy quark mass. Therefore, in this limit  the
 $ \vert uudQ\bar{Q}\rangle $ Fock state probability distribution takes the form
 \begin{eqnarray}\label{E311}
P(x_1,...,x_5)= {\cal N}_5\delta (1-\sum _{i=1}^5 x_i)\frac{x_4^2x_5^2}{(x_4+x_5)^2},
\end{eqnarray}
where $ {\cal N}_5= {\cal N}/m_{Q,\bar{Q}}^4 $
and its value is determined from Eq.~(\ref{p5})  so that
 $ {\cal N}_5=3600 {\cal P}_5^{Q\bar{Q}}$. Finally,
 the probability distribution for the intrinsic heavy  quark
in the proton obtained by integrating over $ dx_1...dx_4  $ is given by
 \begin{eqnarray}\label{heavyBHPS}
P(x_5) & \! =  \!&
   {\cal P}_5^{Q\bar{Q}} 1800~x_5^2
	  \Big[ \frac{(1-x_5)}{3}
	  \left( 1 + 10x_5 + x_5^2 \right)
		+ 2 x_5 (1+x_5) \ln(x_5) \Big].
 \end{eqnarray}

 According to Eq.~(\ref{E3})
  when we use the same value for the mass of $ q $
and $\bar{q}  $, we obtain equal probability
distributions for them in the five-quark state of the proton.
According to BHPS \cite{BHPS1} assumption with 1\% probability for
intrinsic charm (IC) in the proton we have 

\begin{eqnarray}\label{E3c}
c(x)= 18 x^2
	  \Big[ \frac{(1-x)}{3}\left( 1 + 10x + x^2 \right)
		+ 2 x (1+x) \ln(x)
	  \Big],
\end{eqnarray}
where for convenience, we used $ x $ in place of $ x_5 $.
Although an estimation of the order of
1\% for the probability of finding intrinsic charm in the proton have been
found before \cite{MIT,BHPS1}, but
the first detailed global  analysis of PDFs including the intrinsic charm
component were performed by CTEQ \cite{cteq66}. In Ref. \cite{cteq66} the probability
for IC can be 2-3 times larger than previous predictions.
Most recently, two global analyses 
estimated the probability of  IC and they reached different
conclusions about the possibility of IC.
 The first one presented by CTEQ collaboration  
 \cite{Dulat} using CT10 framework \cite{Gribov:1972ri}.
 According to this refrence, a broader possible probability value for IC is 2.5\%.
The second one, presented by Jimenez-Delgado {\it et al.}  \cite{Jimenez-Delgado:2014zga}, 
they found that the value of IC is at most 0.5\%. There are some 
differences between these two QCD analyses. 
In Ref.  \cite{Jimenez-Delgado:2014zga}, they used less restrictive kinematical cuts 
 $ Q^2 \gtrsim 1  $ GeV$ ^2 $ and $ W^2\gtrsim3.5 $ GeV$ ^2 $ in their analysis in contrast 
 to the previous works. 
 These changes in kinematical cuts lead to a large number of SLAC 
 data related to the lower $ Q^2  $ and $ W^2 $, that might be 
 sensitive to the IC contribution, to be included in QCD fit. In 
 addition, in Ref. 
  \cite{Jimenez-Delgado:2014zga}, they have included EMC heavy structure function 
 $  F_2^{c} $ data \cite{EMC} which can be a possible evidence for existence of the intrinsic charm quark.
 Using various data in these two global analyses of PDFs, 
 affected the obtained results of IC probability from these analyses. 
 As we will demonstrate in this paper, D0 data can be another 
 evidence for IC contribution. 
 So including these data in a global QCD fit could
help place additional bounds on IC probability. Although the 
determination of IC contribution must be done in a global QCD fit but, 
with using the discussed  technique in this paper, for investigating the 
impact of IC on physical observables, we can change the amount of
IC contributions without performing
a complete global analysis for each case.  
In the present study, in addition to 1\% IC, we choose the  value of   
3.5\% IC to investigate the impact of
upper limit of IC on the physical observables such as photon production 
in association with a charm quark. However, it should be noted that the actual
value of IC which is more likely lower than 3.5, would be more reasonable.

According to Eq.~(\ref{heavyBHPS}),  the distribution
of the intrinsic bottom (IB) quarks and intrinsic charm quarks are identical.
 But  it is different in the
 value of normalization. The probability for
IB  is expected to be
${\cal P}_5^{b\bar{b}}={\cal P}_5^{c\bar{c}} (m_c^2/m_b^2)\sim0.001  $
(with $ {\cal P}_5^{c\bar{c}}=0.01 $)
where $ m_c\simeq1.3 $~GeV and $ m_b\simeq4.2 $~GeV
are the masses of charm and bottom quarks, respectively.
As the probability of finding IB in the proton is smaller
than IC by a factor of 0.1, the experimental search
for finding an IC signature in  $ pp(\bar{p}) $ collisions is more interested than IB.

To calculate  the intrinsic light quark distributions
 we can not neglect the mass
of the proton and light quarks.
In this regard, we need to compute the Eq.~(\ref{E3}) numerically.
Recently, Chang
and Peng \cite{Chang:2011du}
calculated  the intrinsic quarks distribution using  Monte Carlo techniques and then extracted
 the probabilities for the intrinsic light quark Fock states.
 In this work, we also calculate the intrinsic quark distributions without
 any assumptions to neglect the proton and light
 quarks masses.
 To check our method, for the case of intrinsic charm
 we used $ m_p = m_u = m_d = 0 $
 in our calculation and
we found the result completely equal to Eq.~(\ref{E3c}).
In this way, for the case of intrinsic strange quark, we choose $ m_u=m_d=0.3  $~GeV
 for the mass of $ u $ and $ d $ quarks and $ m_s = 0.5  $~GeV
  for the mass of $ s $ 
  quark. The parametrization form is given by
 \begin{eqnarray}
s(x)
  & \! =  \! & {\cal P}_5^{s\bar{s}}\times  13188.9~ x ^{1.627} (1 - x )^{10.152}
    (0.029+x^{3.713}+x^{7.426}).
	\end{eqnarray}

\section{The evolution of intrinsic quark distributions}\label{sec:evo}
Heavy quarks play  a crucial role in the study of
many  processes
such as single-top production
and Higgs production in the standard model
and beyond which are quite sensitive to the heavy quark content  \cite{pump}.
Furthermore, having a precise knowledge about the heavy
quark components can help to understand
the fundamental structure of the nucleon.
 In the standard global analysis of PDFs, heavy quarks
 distribution are assumed zero for  $ Q^2 < m_Q^2 $. In this way,
 we do not need any parameterization form  for the heavy quark distributions
 and  they arise perturbatively through the splitting of gluons
 into quark-antiquark pairs in  DGLAP  $ Q^2 $-evolution equations.
 Moreover, in these analysis it is usually assumed that
 there is no intrinsic heavy quark IQ
 in the proton.
Although there is no theoretical reason to reject this assumption,
there are some data that suggest the existence of
an intrinsic heavy component in the proton \cite{EMC}.
In addition to the experimental evidence,  
in the light-cone picture of the proton the existence
of the intrinsic components in the proton wave function is
inevitable; these kinds of quarks are certainly
  non-perturbative in origin
 and can play an important role at  high $ x $.
 Therefore, in order to investigate  the impact of
 intrinsic heavy quarks on the physical observables,
 the study of the evolution of their distributions together
with extrinsic ones  can be interesting and useful study.

Since the intrinsic heavy quark distribution
gives an insignificant contribution to the evolution
of light quark and gluon distributions,
one can use the standard approach for global analysis of PDFs
without considering an intrinsic heavy quark component
in gluon and light quark distributions.

The DGLAP evolution equation  that   used
for the standard approach  of global analysis has compact form as \cite{Altarelli:1977zs,Gribov:1972ri,Dokshitzer:1977sg}
\begin{equation}
\dot f_i = \sum_{j=q,g,Q}P_{ij}\otimes f_j \, . 
\end{equation}

For the heavy quark distribution, the scenario is different. 
 If we adopt the intrinsic quarks in the proton,
 then the total heavy quark distribution in any $ x $ and
$  Q^2 $ region can be obtained by adding the intrinsic
contribution (non-perturbative) $ xQ_{int} $ to
the extrinsic component (perturbative) $ xQ_{ext} $ as follows
\begin{equation}\label{tot}
xQ(x,Q^2)=xQ_{ext}(x,Q^2)+xQ_{int}(x,Q^2),
\end{equation}
where $ Q=c, b $ and we use the short-hand notation
$ xQ(x,Q^2)\equiv xf_Q(x,Q^2) $. It should be noted that it is not possible to mix these two terms in the boundary conditions for
the QCD evolution.
In this case,
 the evolution equation of heavy quarks can be separated
into two independent parts.
The first part is evolution of the extrinsic heavy quark. 
The PDFs for  the extrinsic heavy  component, like gluon 
and light quark,  can be taken from a global analysis
 result of various groups providing global analysis of 
 PDFs \cite{CTEQ,MSTW,NNPDF,DelDebbio:2007ee,ABM,JR,KKT},
 for example CTEQ66 \cite{cteq66}
  which are
 available in the Les
Houches Accord PDF Interface (LHAPDF \cite{Bourilkov:2006cj}) in arbitrary  $ x $ and $ Q^2 $.
The second part is evolution of the intrinsic heavy quark 
distribution $  Q_{int}$.
The $ Q^2 $-evolution of the intrinsic heavy
quark distribution is controlled by non-singlet evolution equation.
 According to Ref.\cite{Lyonnet:2015dca}, non-singlet evolution provides a good approximation for
  evolution of the intrinsic heavy quark distributions as
\begin{equation}
\dot Q_{int} = P_{QQ}\otimes Q_{int}  . 
\end{equation}

Therefore,
non-singlet technique allows us
to evolve intrinsic heavy quark distribution independently from the gluon and
other PDFs.
We should care the momentum sum rule for all PDFs in a global QCD analysis, if we take into account the above non-singlet evolution equation for intrinsic heavy quarks.
But as discussed in Ref.\cite{Lyonnet:2015dca}, we have very small violation of this sum rule for intrinsic heavy quarks.

 In recent years the CTEQ
collaboration has done some global analysis of PDFs
considering intrinsic charm from
the BHPS model  \cite{BHPS1}
by adding its contribution for $ Q\geqslant m_c $
to the charm content of the proton and presented
CTEQ66c PDF sets \cite{cteq66} for intrinsic quark.
By using their results, we can  only have
the total charm distribution  in any  $ x $ and
$  Q^2 $
but we do not have the intrinsic contribution separately.
whereas non-singlet evolution of the intrinsic heavy quark component 
of the parton allows us to study the impact of 
 this nonperturbative contribution on the physical observable 
 without performing a complete global analysis of PDFs.
 In other word, this technique gives us evolution of
 the intrinsic charm distribution
 in any $ x $ and $ Q^2 $ and can  be added to any PDFs.
We carried out our calculation by QCDNUM
 package \cite{QCDNUM} and used its ability for the evolution
 of the non-singlet PDFs.

   \begin{figure}[t!]
\centering
    \includegraphics[width=1.01\textwidth,clip]{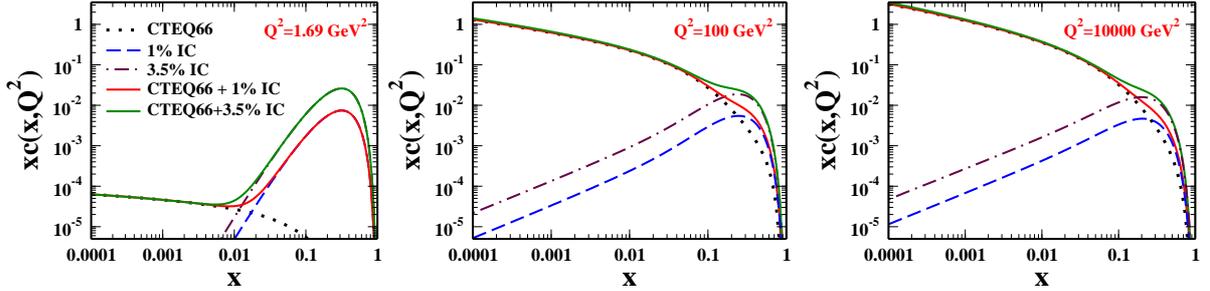}
 \caption{Distributions of the charm quark in the proton.
 The dotted  curve shows the extrinsic
charm distribution from CTEQ66 at $ Q^2=1.69, 100 $
 and $ 10000 $~GeV$^2$ \cite{cteq66}. The dashed blue and dotted-dashed  maroon curves
correspond to our results for the 1\% and 3.5\% intrinsic charm distribution at mentioned $Q^2$ values.
The solid red curve  displays the total charm distribution
considering a 1\% IC  from  our grid plus CTEQ66 PDFs and 
the solid green curve  presents  the total charm distribution
considering a 3.5\% IC  from  our grid plus CTEQ66 PDFs.
}\label{intrinsic}
\end{figure}

 Fig.~\ref{intrinsic}  shows the
$ x $ distribution of
extrinsic charm from the CETQ66 \cite{cteq66}
and intrinsic charm with 1\% and 3.5\% probability
by the method we described above
at $ Q^2=1.69,100 $ and $ 10000 $~GeV$^2$.
In this figure, we also present  the total
charm distribution which is the sum of
extrinsic and intrinsic components (1\% and 3.5\%).
As can be seen, the intrinsic distribution
starts from zero
  and its behaviour is like the valence distributions
  so that if we integrate it, we can obtain  ${\cal P}_5^{c\bar{c}}$.
To check the evolution, one can extract the extrinsic charm distribution at fixed $Q^2$
using CTEQ66 PDFs and then add it to  IC contribution using our grids to compare this total charm distribution
with extracted results from CTEQ66c
 \cite{cteq66}.
 This comparison shows that,
  there is a good agreement between our result
 and CTEQ66c.
 As a result of the evolution of intrinsic distribution,
one can see that the intrinsic distribution's peak decreases in magnitude
and also shifts to the smaller values of  $ x $ just like
the valence quark behaviour as expected.
The  $ Q^2 $-evolution of the extrinsic charm  distribution
 is dominated at small $ x $
 and have a sea like behaviour so that at higher $ Q^2 $,
 its magnitude  increases at small $ x $  in the DGLAP evolution regime.
 Thus, the  $ Q^2 $-evolution of the total charm
 distribution in the proton containing extrinsic contribution
  from CETQ66 \cite{cteq66} and intrinsic ones from the
  BHPS \cite{BHPS1}  increases at small $ x $ and decreases at high $ x $ at higher $ Q^2 $.
Our results are compatible with very recently reported results in Ref.~ \cite{Lyonnet:2015dca}.
It should be noted that we use NLO extrinsic PDFs from CTEQ66 sets \cite{cteq66} 
  and our calculations for IC non-singlet evolution are preformed 
 at the NLO approximation as well.

 Actually, the significant difference between our work and CTEQ66c is
 that, in our work
there is no limitation to choose the value of    ${\cal P}_5^{c\bar{c}}$.
Thus, one can have the $ x $ distribution of intrinsic charm
for any ${\cal P}_5^{c\bar{c}}$ at arbitrary  $ Q^2 $.

  Since having the evolution of intrinsic quark distributions
  are important at large momentum fraction $x$,
  we prepare a set of grid files for the intrinsic strange, charm and bottom distributions
   and also an interpolation
  code to evolute these distributions to
  any $ x $ and $ Q^2 $ values \cite{grid}. It should be noted that, one can
  choose and fix the value for the probability
  of intrinsic quark ${\cal P}_5^{q\bar{q}}$ as an input parameter.

\section{ The prompt  photon production in association with a c-jet}\label{sec:pho}
\begin{figure}[h!]
\centering
    \includegraphics[width=0.7\textwidth,clip]{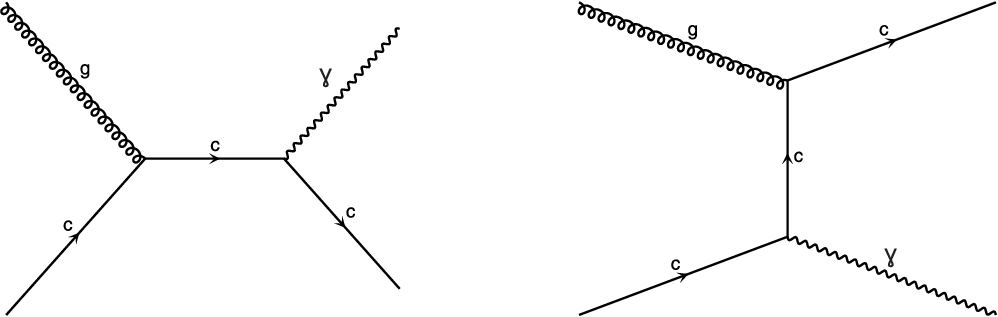}
 \caption{ The Feynman diagrams for the QCD leading order
 contribution of the Compton process $ g c \rightarrow \gamma c $
  in the s-channel (left) and in the t-channel (right). }\label{Fig2}
\end{figure}

The prompt photon production in association with charm
quark jet process can provide some information of parton
distribution functions and  improve our
understanding of the perturbative techniques applied to
calculate the hard scattering sub-process and also
investigate the possibility of intrinsic charm quark
component in the proton at larger $ x $ \cite{Bednyakov}.
According to Fig.~\ref{Fig2}, at the leading order (LO), 
the main contribution arises from the Compton sub-process
$ g c \rightarrow \gamma c $. Within the leading order, the inclusive $ \gamma+c  $ production can 
also  originate from the sub-processes $ g g \rightarrow c \bar{c} \gamma$,
  $ c g \rightarrow c g \gamma$ or $ q c \rightarrow q c  \gamma$
  where  the fragmentation    of quarks or gluons produces
 a photon. 

At the next-to-leading order (NLO), the number of contributing sub-processes
increases.  In this way, the photon component includes
contributions from  $ gg \rightarrow  \gamma c\bar{c}$,
$ g c \rightarrow  \gamma g c $,
$ c q \rightarrow  \gamma  q c$,
$ c \bar{q} \rightarrow  \gamma  \bar{q} c$,
$ c\bar{c} \rightarrow  \gamma c\bar{c}$,
$ cc  \rightarrow  \gamma cc$
and  the annihilation sub-process $ q\bar{q}\rightarrow  \gamma c\bar{c}$ \cite{Stavreva,Stavreva1} ,
which apart from $ q\bar{q}\rightarrow  \gamma c\bar{c}$,
all sub-processes are $ g $ and $ c $ PDF initiated.
 At the LHC, the Compton process dominates
for all energies, whereas at the Tevatron  the annihilation process
 $ q\bar{q}\rightarrow  \gamma c\bar{c}$
 dominates for photons with high transverse momentum $ p_T^\gamma $ \cite{Stavreva}.

\subsection{Comparison to Tevatron data}
Recent years, the prompt photon and heavy quark jet
production in $ p\bar{p} $ collisions at the Tevatron
have been investigated \cite{D0,Abazov:2009de,Aaltonen:2009wc,Abazov:2012ea};
this process can be very useful
for testing the possible existence of intrinsic quarks in the nucleon.
For example, in Ref.~\cite{D0} the differential cross section for the
associated production of a $ c $-quark jet and an isolated photon
with rapidity $\vert y^\gamma \vert < 1.0  $ and transverse
momentum  $ 30 < p^\gamma_T<300 $~GeV have been
measured as a function of $ p_T^\gamma $ at  $ \sqrt{s}=1.96$~TeV so that
 the $  c $-jet  has  $\vert \eta^{c} \vert < 1.5  $ and  $  p^c_T>15 $~GeV.

\begin{figure}[h!]
\centering
 \includegraphics[width=0.65\textwidth,clip]{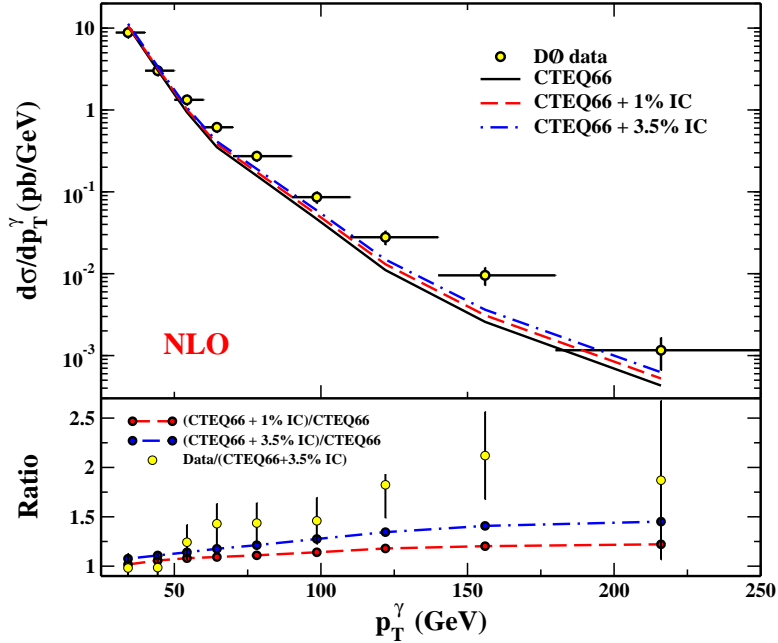}
\caption{ A comparison of D0 measurement of differential
$ \gamma+c $-jet cross section as a function of
$ p_T^\gamma $  at $ \sqrt{s}=1.96$~TeV with $\vert \eta^{c} \vert < 1.5  $
and $\vert y^\gamma \vert < 1.0  $  \cite{D0}
and corresponding the NLO theoretical calculations.
The solid  curves are calculated using
CTEQ66 (pure extrinsic) \cite{cteq66}.
The dashed and dotted-dashed curves
   correspond to the inclusion of 1\% and 3.5\% IC
  from the BHPS, respectively and using our grids files \cite{grid}(top).
 The ratio of these spectra (including IC contribution) to the pure extrinsic CTEQ66 is 
 shown in the bottom panel. The ratio of $ \gamma+c $-jet cross sections
 for data to CTEQ66 plus 3.5\% IC presented by yellow points.
}
\label{D0}
\end{figure}

Fig.~\ref{D0} shows a comparison of the D0
measurement of the differential $ \gamma+c $-jet cross section as a
 function of $ p_T^\gamma $ \cite{D0}
and the corresponding NLO theoretical calculation. 
This calculations  have been carried out by the MadGraph \cite{madgraph}.
The lowest curve is related to CTEQ66 PDFs \cite{cteq66}
without the IC contribution and as it is clear, it has a poor
description of data. At large $ p_T^\gamma $ region,
as expected, the spectrum grows by the inclusion of the
IC contribution. 
  In this figure the solid, dashed and    dotted-dashed  curves
 represent the theoretical results for the cross section
 using the CTEQ66 \cite{cteq66} (without IC contribution) and
CTEQ66 plus  BHPS with 1\% and 3.5\% IC, respectively.
One can see  the obtained result considering 3.5\% IC has
 the better description of data.
   In the bottom of
 Fig.~\ref{D0}, the ratio of  CTEQ66 PDFs
  adding  1\% and 3.5\% IC to CTEQ66 PDFs is illustrated.
 This ratio for 3.5\% IC is about 1.5
 when   $ p_T^\gamma$ reaches 216~GeV, while 
 this factor for a 1\% IC contribution is about 1.2.
 Also, in the bottom of Fig.~\ref{D0}, we showed the ration 
 of data to the result of CTEQ66 plus 3.5\% IC by yellow points.

Although, we could 
also use the CTEQ66c PDFs
but as mentioned in previous section,
we can choose any value for  ${\cal P}_5^{c\bar{c}}$ in our calculation. 
Using  CTEQ66c PDFs and our results in this calculation are in good agreement with each other.

\subsection{Predictions for the LHC}
The LHC with $  pp $-collisions, operate at the
center of mass energy, $ \sqrt{s}=7-14$~TeV, which is
much greater than the Tevatron.
In order to make prediction of the inclusive production of
$ \gamma+c $-jet process at the LHC, we need to select
kinematical regions where are the most
sensitive to the intrinsic charm quark contribution.
To this, we have used the kinematical regions which analyzed in detail by
 V.~A.~Bednyakov~{\it et al.} \cite{Bednyakov}.

 The differential $ \gamma+c $-jet cross section in $ pp $ collisions versus the transverse
 momentum of the photon is presented
 for the photon rapidity  $1.52<\vert y^\gamma \vert < 2.37  $
 at $ \sqrt{s}=8  $~TeV and for
 transverse momentum $ 50 < p^\gamma_T<400 $~GeV.
The $  c $-jet also has  $\vert \eta^{c} \vert < 2.4  $
 and  $  p^c_T>20 $~GeV.
In this kinematical region, the charm momentum fraction is larger than
0.1 ($ x_c >0.1 $) where the  intrinsic charm distribution
is completely considerable in comparison with the extrinsic  charm distribution.

\begin{figure}[h!]
\centering
\includegraphics[width=0.65\textwidth,clip]{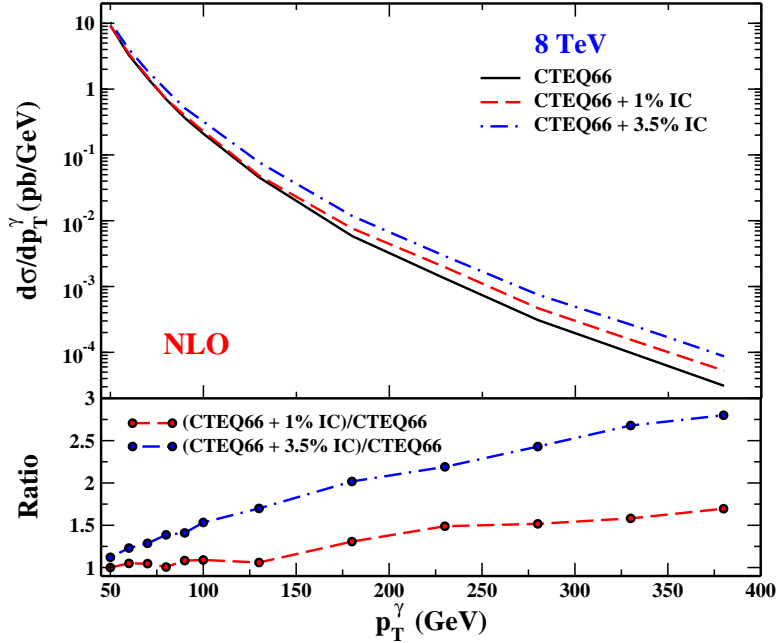}
\caption{  The differential $ \gamma+c $-jet cross section in
$ pp $ collisions as a function of $ p_T^\gamma $ for  NLO theoretical calculations at 
$ \sqrt{s}=8$~TeV and using CTEQ66 without IC contribution (solid curve),
CTEQ66 \cite{cteq66} plus 1\%
(dashed curve) and 3.5\%
(dotted-dashed curve) IC from grid files \cite{grid} (top).
The ratio of these spectra (including IC contribution) to the pure extrinsic CTEQ66  is 
 shown in the bottom panel.
}
\label{LHC}
\end{figure}

In Fig.~\ref{LHC} the differential cross section
 $ d\sigma/dp_T^\gamma $ for $ pp \rightarrow  \gamma+c $-jet process
calculated at NLO is presented as a function of
the transverse momentum of photon.
In this kinematic region,  considering IC  contribution  can substantially increase
 the cross section values especially at high $ p_T^\gamma$.
 In this figure we
 present the  results using CTEQ66 and
 CTEQ66 plus 1\% IC.
  Moreover for the comparison, we present the results obtained using
  CTEQ66   plus 3.5\% IC.
 The difference between the results is clearly visible in bottom of
 Fig.~\ref{LHC} where the ratio of the spectra including
 IC contribution with 1\% and 3.5\% IC  probability to CTEQ66
 is presented as a function of  $ p_T^\gamma$.
 By comparing the results, one can recognize that the values of the spectra
increases considering IC contribution so that the BHPS with 3.5\% IC result
is placed above the CTEQ66 and CTEQ66 plus 1\% IC.
For example, the inclusion of the 3.5\% IC increases the spectrum
by a factor of 2.8 at $ p_T^\gamma=380$, while for the 1\% IC  this factor is
about 1.6.
Our results for
the differential cross section
of $ pp \rightarrow  \gamma+c $-jet process
 at $ \sqrt{s}=8$~TeV
are in good agreement with recent published results \cite{Bednyakov}.

To  further  investigate  the role of intrinsic charm
in the proton, we give predictions for the
LHC, but at the
center of mass energy 
$ \sqrt{s}=13$~TeV.
Therefore, we  used the kinematical regions which is
presented  in Ref.~\cite{Bednyakov}.
According to Ref.~ \cite{Bednyakov} we have
\begin{equation}\label{xf} 
x_c\geq x_F = \frac{2p_T}{\sqrt{s}}\sinh (\eta),
\end{equation}
where $  x_F  $ is the Feynman  scaling variable   of the produced hadron
and $ x_c $
is scaling variable of the intrinsic charm quark in the proton.
As well as, $ \eta $ is pseudo rapidity of photon and $ p_T $ is its transverse momentum.
According to presented kinematic cut  at $ \sqrt{s}=8$~TeV
by  V.~A.~Bednyakov~et al. \cite{Bednyakov}
for $ x_c\geq 0.1 $,  the values of
the photon transverse momentum $ p_T$ will be changed
 by a factor $ 13/8 $  for the photon rapidity  $1.52<\vert y^\gamma \vert < 2.37  $
and
 transverse momentum $ 50 < p^\gamma_T<400 $~GeV at $ \sqrt{s}=13$~TeV.

   \begin{figure} [ht]
\begin{center}
\includegraphics[width=0.65\textwidth,clip]{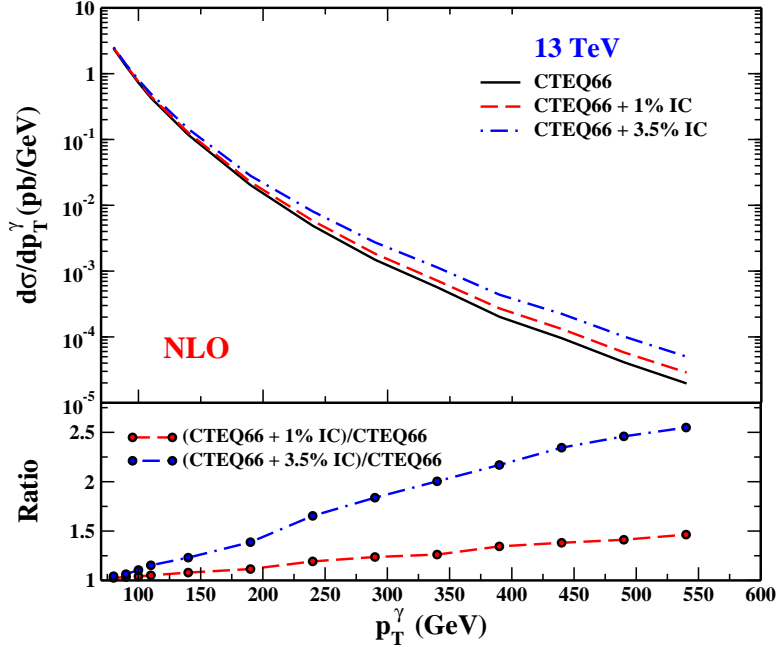}
\caption{The differential $ \gamma+c $-jet cross section in
$ pp $ collisions as a function of $ p_T^\gamma $ at the NLO and
$ \sqrt{s}=13$~TeV and using CTEQ66 without IC contribution (solid curve),
CTEQ66  plus 1\% IC  (dashed curve),
CTEQ66  plus 3.5\% IC  (dotted-dashed   curve),
for forward photon rapidity $1.52<\vert y^\gamma \vert < 2.37 $ (top).
The ratio of these spectra for 1\% (dashed curve) and 3.5\% (solid curve) IC contribution  to CTEQ66  is 
 shown in the bottom panel.
}\label{LHCNLO13tev}
\end{center}
\end{figure}

The differential cross section
 $ d\sigma/dp_T^\gamma $ for $ pp \rightarrow  \gamma+c $-jet process
  calculated at NLO  as a function of
 transverse momentum of the photon has been shown in  the Fig.\ref{LHCNLO13tev}.
Here, we choose the rapidity of photon and $c$-jet,  $1.52<\vert y^\gamma \vert < 2.37  $
and $\vert \eta^{c} \vert < 2.4  $, respectively. We use also the  $  p^c_T>20 $~GeV and $ 80 < p^\gamma_T<540 $~GeV for c-jet and photon transverse momentum.

Similar way to our result for  $ \sqrt{s}=8$~TeV,  the IC  contribution  can substantially increase
 the cross section values specially at high $ p_T^\gamma$ at $ \sqrt{s}=13$~TeV.
  The   solid and   dashed curves
 represent the calculation results using CTEQ66 and
 CTEQ66 plus 1\% IC,   respectively.
  Moreover, the result obtained using
  CTEQ66   plus 3.5\% IC from the BHPS  presented by dotted-dashed line.
 The difference between the results is clearly visible in bottom of
 Fig.~\ref{LHCNLO13tev} where the ratio of the spectra including
 IC contribution with 1 and 3.5\% IC  probability and
 without it, is presented as a function of  $ p_T^\gamma$.
One can see the values of the spectra
increase considering IC contribution.
In this way, the inclusion of the 3.5\% IC increases the spectrum
by a factor 2.17 at $ p_T^\gamma=380$~GeV, while for the 1\% IC  this factor is
about 1.34. Also this factor at $ p_T^\gamma=540$~GeV for 1 and  3.5\% IC is about 1.46 and 2.45, respectively.

According to Eq. \ref{xf}, at the specific values 
of the pseudo rapidity and the transverse momentum of the  photon, 
by increasing  the IC, $ \sqrt{s} $  decreases. 
For example, at $ p_T^\gamma=280  $ GeV the spectrum increases by a 
factor 1.51 for $ \sqrt{s}=8 $ TeV and 1.21 for $ \sqrt{s}=13 $ TeV at the LHC.
In addition, we should be noted that at 
 the higher rapidity  with a lower
center-of-mass energy, where $ x_c $ would be larger and so the IC 
 contribution in the proton is increased would be preferably
suited to searching and discover  intrinsic charm.
Finally, we should mentioned that the cross section result using 
CT10  PDFs are are very similar to the result of CTEQ66,  
but these values for MSTW are slightly higher than CTEQ66.

\section{Conclusions}\label{sam}
In this work,
we present the evolution of intrinsic charm  distribution which is practical
for intrinsic bottom and strange quark.
An important aspect of our calculation is that we present 
$ xc_{int} $as a function of $ Q^2 $ for arbitrary $ {\cal P}_5^{c\bar{c}} $.
The grid files for the evolution of intrinsic strange, charm and bottom quarks for arbitrary $ {\cal P}_5^{q\bar{q}} $ that were used in this paper are available
in Ref.~ \cite{grid}.
We have presented a comparative
analysis to investigate the role of intrinsic
charm in the results of the inclusive production
of a prompt photon and $ c $-jet in hadron colliders
for two value of $ {\cal P}_5^{c\bar{c}} $.
The calculations were done for the Tevatron
$  p\bar{p} $-collisions at rapidity
$\vert y^\gamma \vert < 1 $, $\vert \eta^c \vert < 1.5  $ and
LHC
$  pp $-collisions at rapidity
 $1.52<\vert y^\gamma \vert < 2.37  $ and
 $\vert \eta^c \vert < 2.4  $ at  $ \sqrt{s}=8$~TeV   and  $ \sqrt{s}=13$~TeV.
 As a result we found, regardless of the value of intrinsic charm probability
 in the proton,
the IC contribution increases the magnitude of the cross section,
and has significant contribution in cross section when the photon transverse momentum grows.
 Finally, we have found that the BHPS with 3.5\% IC
has better compatible to the D0 data than 1\% IC. Nevertheless, to determine
the probability of the intrinsic  charm component in the proton, 
a global analysis  using D0 and LHC data
(which are particularly sensitive to the charm quark)  is required.
We expect that using these data will give a
better understanding of the IC components in the proton.

\section{Acknowledgments}
\vspace{-0.2cm}We thank S. J. Brodsky and F. Olness for reading manuscript and useful discussions and comments. S. R. appreciates O. Mattelaer, G. Lykasov
and V. Bednyakov for discussion and help. A. K.
thanks SITP (Stanford Institute for Theoretical Physics) and the Physics Department of SMU (Southern Methodist
University) for their hospitality at the beginning of this
work. He is also grateful to CERN TH-PH division for the hospitality
where a portion of this work was performed. 

\appendix 
\section{Fortran code}
A \texttt{FORTRAN} package containing the distribution
of intrinsic charm (IC) and bottom (IB) distributions  and
also intrinsic strange (IS) distribution for arbitrary $ x $ and $ Q^2 $  can be found in
\texttt{http://particles.ipm.ir/links/QCD.htm} \cite{grid} or
obtained via e-mail from the authors. It should be noted that,
this code is available for  arbitrary $ {\cal P}_5^{q\bar{q}} $.
 The package includes an example
program to illustrate the use of the routines.

\end{document}